\documentclass{article}

\usepackage{PRIMEarxiv}

\usepackage[utf8]{inputenc} 
\usepackage[T1]{fontenc}    
\usepackage{hyperref}       
\usepackage{url}            
\usepackage{booktabs}       
\usepackage{amsfonts}       
\usepackage{nicefrac}       
\usepackage{microtype}      
\usepackage{lipsum}
\usepackage{fancyhdr}       
\usepackage{graphicx}       
\usepackage{subcaption}
\usepackage{caption}
\usepackage{array}
\usepackage{longtable}

\newcolumntype{P}[1]{>{\raggedright\arraybackslash}p{#1}}
\graphicspath{{media/}}     

\title{Detecting Agitation Before Behavioral Escalation in Autistic Youth Through Multimodal Wearable Sensing
}

\author{
  Nibraas Khan$^{1}$\thanks{Corresponding author: \texttt{nibraas.a.khan@vanderbilt.edu}} \quad
  Abigale Plunk$^{2}$ \quad
  John Staubitz$^{3}$ \quad
  Ingrid Shragge$^{3}$ \\
  \bfseries Jordan Brooks$^{3}$ \quad
  Suzanne Wright$^{3}$ \quad
  Alec Brewer$^{4}$ \quad
  James Dieffenderfer$^{4}$ \\
  \bfseries Alper Bozkurt$^{4}$ \quad
  Amy Weitlauf$^{3}$ \quad
  Nilanjan Sarkar$^{1,2,5}$ \\[1em]
  \small $^{1}$Department of Computer Science, Vanderbilt University, Nashville, TN, USA \\
  \small $^{2}$Department of Electrical and Computer Engineering, Vanderbilt University, Nashville, TN, USA \\
  \small $^{3}$Treatment and Research Institute for Autism Spectrum Disorders (TRIAD), \\
  \small Vanderbilt University Medical Center, Nashville, TN, USA \\
  \small $^{4}$Department of Electrical and Computer Engineering, North Carolina State University, Raleigh, NC, USA \\
  \small $^{5}$Department of Mechanical Engineering, Vanderbilt University, Nashville, TN, USA
}

\begin{document}
\maketitle

\begin{abstract}
Challenging behaviors including aggression, self-injury, and property destruction are observed in 68\% of autistic youth and pose risks to youth and caregivers. These episodes are preceded by agitation, a rising state of distress expressed through movement, vocalization, and autonomic arousal. Its signs are subtle and individualized, and its autonomic components are invisible without instrumentation. We collected upper-body movement from inertial measurement units, physiology from a wrist-worn device, and vocalizations from lapel microphones across 30 clinician-led sessions with 15 autistic youth, paired with expert behavioral annotations. We adapt four pretrained foundation models, one per modality, project each to a shared 128-dimensional space, and fuse them into a single group model. The model detected agitation with an area under the ROC curve of $0.724$ at the clinician-annotated onset (within-participant permutation $p=0.0005$), declining to $0.608$ at $30$\,s before onset. Thirteen of fifteen participants were above chance. A from-scratch configuration reached only $0.58$, while frozen and fine-tuned features performed comparably ($0.71$ and $0.72$). Audio contributed most of the signal, and a watch-only configuration stayed near chance. Individualized agitation is therefore detectable, including in unannotated windows preceding the annotated onset, using foundation-model transfer with one shared model rather than one per child.
\end{abstract}


\section{Introduction}

\label{sec:introduction}

Autism spectrum disorder (ASD) is a heterogeneous neurodevelopmental condition characterized by differences in social communication and social interaction, along with restricted, repetitive patterns of behavior, interests, and activities \cite{cdcDSM}. Globally, the World Health Organization estimates that 1 in 100 children have ASD \cite{whoAutism}, while in the United States, a 2022 CDC survey identified approximately 1 in 31 eight-year-old children with ASD \cite{cdcADDM2025}. Among the most pressing clinical concerns in ASD are challenging behaviors including aggression, self-injury, and property destruction. Across childhood and adolescence, as many as 68\% of autistic youth exhibit such behaviors \cite{kanne2011aggression}. Persistent episodes pose risks to youth and caregivers alike \cite{chadwick2000factors}, impede skill acquisition \cite{emerson2001challenging}, and can lead to exclusion from school services and community opportunities, diminishing quality of life \cite{parish2012state}. These behaviors often require intensive interventions including specialized behavioral therapy, crisis response, caregiver training, and school-based supports \cite{jeglum2024emergency}. Understanding why and under which circumstances these behaviors occur is therefore critical to effective and safe intervention.

Because these behaviors can create unsafe situations, clinicians seek to understand their function, the pattern of environmental variables that evoke and reinforce them, and intervene preventively. Applied Behavior Analysis (ABA) provides an evidence-based framework for this process. It is implemented by Board Certified Behavior Analysts (BCBAs) in collaboration with caregivers and service systems \cite{foxx2008applied}. BCBAs draw from a continuum of assessment methods, ranging from indirect tools such as structured interviews and rating scales, to direct observation, to experimental analyses that test causal hypotheses about behavioral function \cite{fee2016agreement,o2010functional,slaton2017interview,iovino2022new,fisher2016comparisons}.

A common technique is the Interview-Informed Synthesized Contingency Analysis (IISCA; \cite{slaton2017interview}), a brief analytic procedure within the broader category of Practical Functional Assessment (PFA; \cite{hanley2003functional}). In an IISCA, a clinician first interviews caregivers to identify the conditions under which challenging behavior is likely and the individualized signs that precede it, then arranges those conditions in a controlled session and withdraws them as soon as those signs appear. The session ends the escalation rather than allowing it to run to a high-intensity episode.

Those signs are the target of this work. We refer to the escalating state they express as \emph{agitation}: a condition of rising distress expressed through changes in various modalities such as movement, vocalization, and autonomic arousal \cite{khan2024pilot}. Agitation is not a single discrete act; rather, it builds over seconds and unfolds across multiple expressive and physiological channels that differ in how readily a person can observe them. The expressive channels comprise the visible and audible manifestations of distress, such as an abrupt shift in posture, the tightening of muscles in the jaw or shoulders, a sudden change in vocal pitch, or vocal protests like heavy breathing or pacing. Caregivers and clinicians become highly skilled at recognizing these external, idiosyncratic behavioral signs. However, agitation is simultaneously driven by internal autonomic channels. These unobservable manifestations include sympathetic nervous system responses like a surging heart rate, diminishing heart rate variability, and spikes in electrodermal activity. Because these physiological components occur entirely beneath the skin, they remain invisible to even the most attentive observer without specialized instrumentation \cite{mccarty2016fight,critchley2002electrodermal}.

However, noting and recording these behaviors manually at scale can be technically challenging and resource-intensive. BCBAs are in high demand, services are costly, and dedicated human observation and data collection may compete with providing intervention, instruction, or care in schools, clinics, and homes. Many early signs of agitation are subtle (posture shifts, micro-gestures, brief vocalizations) and easy to miss without continuous attention, and its autonomic components cannot be seen at all. Coverage is inherently incomplete across settings and times of day, observer drift and reactivity can erode data quality, and privacy constraints on recording minors further limit where and how long continuous observation can occur.

These constraints motivate technologies that render the signs of behavioral escalation measurable and scalable. Video analysis can quantify posture and repetitive movement \cite{barami2024automated}, but camera-based approaches are constrained by occlusion, lighting, and privacy \cite{de2020computer}. Body-worn Inertial Measurement Units (IMUs) extend coverage beyond the camera's view \cite{cantin2020detecting,albinali2009recognizing,goodwin2019predicting}, yet both video and IMU-based tools operate only once something is externally visible. Escalation often begins earlier in the autonomic nervous system, where Heart Rate (HR) increases, heart-rate variability decreases, and electrodermal activity rises \cite{mccarty2016fight,taelman2009influence,critchley2002electrodermal}. Across sensing modalities, however, most studies train models to predict the challenging behavior itself rather than the agitation that precedes it. Targeting agitation directly creates a safer window for least-restrictive, function-based supports, because the state is present and measurable before the episode it leads to. It also improves data quality: teams can label reliable, lower-intensity signals without waiting for high-intensity episodes, increasing the number of positive training windows \cite{heath2019precursor}.

Prior work established the feasibility of treating agitation as the supervised target for real-time early warning in clinical settings, using a lightweight multimodal sensor suite \cite{khan2024pilot}. In that pilot study with three participants, a multimodal wearable stack spanning movement, peripheral physiology, and audio with AdaBoost achieved an average recall of 69.97\% at labeled onset and 55.81\% at 25\,s prior, demonstrating that meaningful early warning is achievable in practice \cite{khan2024pilot}. However, while the pilot study demonstrated feasibility, it highlighted critical bottlenecks for clinical scaling. These included a constrained sensor suite and a reliance on models trained entirely from scratch on a small number of labeled events. Because agitation is highly idiosyncratic, capturing it effectively requires robust and comfortable sensing deployed across a larger population, alongside methods that overcome severe data scarcity without losing sensitivity to individual differences. Furthermore, attempting to predict highly specific behavioral signs spreads limited data too thinly as the diversity of those signs increases. We address that scarcity by moving the representation outside the cohort. Foundation models pretrained on movement, audio, photoplethysmography, and electrodermal corpora supply general structure for each signal type, learned from populations and tasks unrelated to agitation. Our labeled events are then spent on the decision boundary, and not on learning what a vocalization or an arm movement looks like in the first place. The present work targets these barriers with four primary contributions:

\begin{itemize}
    \item \textbf{Clinically grounded multimodal dataset:} We scaled collection to 15 participants across 30 sessions using an improved sensor array informed by a formative wearability study. The dataset captures upper body movement, physiology, and vocalizations during clinical ABA sessions, paired with BCBA verified annotations.
    \item \textbf{Agitation modeling via foundation models:} We overcome data scarcity and label fragmentation by pooling individualized signs into a single agitation class and fusing pretrained foundation models into a shared representation.
    \item \textbf{Evaluation:} We evaluate encoder transfer (frozen, fine tuned, and from scratch), quantify modality contributions via ablation, and demonstrate that a shared group model performs on par with individual per participant models. We report all results using area under the ROC curve with bootstrap confidence intervals and within participant permutation tests.
    \item \textbf{Analysis of cohort heterogeneity:} We analyze why detection succeeds or fails across different children. By mapping model performance to the specific phenotypic expression of a child's agitation, such as the difference between vocal signs and diffuse postural signs, we provide interpretable boundaries for future clinical deployment.
\end{itemize}

The rest of the paper is structured as follows: Section~\ref{sec:related_works} reviews related work on sensing for challenging behaviors and for the agitation that precedes them. Section~\ref{sec:methods} describes the dataset, sensing modalities, feature design, and model. Section~\ref{sec:results} presents quantitative results and model interpretability analyses. Section~\ref{sec:discussion} offers discussion of deployment, ethics, and limitations. Section~\ref{sec:conclusion} concludes and outlines future work.

\section{Related Work}
\label{sec:related_works}

Sensing has been applied to challenging behaviors and to the earlier, lower intensity states that precede them; yet much of the literature still targets the challenging behaviors themselves rather than explicitly modeling agitation. To prepare the reader for systems that act early in real settings, we frame the related work by the channels through which agitation is expressed: what can be seen and heard (expressive signals, meaning movement and vocalizations, captured by vision, body worn IMUs, and audio) and what reflects internal state (autonomic signals measured by physiology). We then discuss how prior multimodal systems have attempted to connect these channels, and highlight how leveraging pretrained representations addresses the fundamental data constraints of personalized detection.

In ABA practice, BCBAs begin with what can be seen: direct observation of antecedents, behaviors, and consequences. It is therefore unsurprising that early computational systems mirrored this visual emphasis, adapting video pipelines to quantify visible topographies. A systematic review highlighted the promise and pitfalls of computer vision in autism research, underscoring the importance of robust pose estimation and ecological validity \cite{de2020computer}. Recent work has scaled to longer, more naturalistic recordings: an open-source pipeline automatically localized stereotyped motor movements in hours of clinical video from 241 children with high segment-retrieval sensitivity \cite{barami2024automated}. Targeted detectors also show strong accuracy when tasks are narrowly defined; for example, Washington et al.\ achieved a macro-F1 of $90.8\%$ for head-banging classification using head-pose keypoints with a CNN+LSTM under child-wise splits \cite{washington2021activity}. Despite these advances, real-world deployment must contend with occlusion, variable lighting, bystander privacy, and the fact that much of the day occurs outside any camera's view \cite{de2020computer,barami2024automated}. These constraints motivate body-worn sensing.

Body-worn inertial sensors can capture posture tightening, abrupt limb actions, repetitive movements, and other motor expressions of agitation and challenging behavior. Early lab/classroom work recognized stereotypy from accelerometry with $\sim$89\% accuracy and classroom per-class F1 of 0.74--0.94 \cite{albinali2009recognizing}. Subsequent studies extended to naturalistic self-injurious behavior, reporting up to 99.1\% individual-level accuracy \cite{cantin2020detecting}, while design studies emphasized garment-integrated multi-IMU arrays for comfort and adherence in schools \cite{scheithauer2022feasibility}. Yet, motor changes are not the sole expressive channel during escalation as brief vocal signals often co-occur and can add complementary evidence.

Microphones can capture protests, non-speech vocalizations, and prosodic shifts that accompany agitation, especially in busy classrooms where ambient mics underperform. Meta-analytic and cumulative work documents robust (though heterogeneous) acoustic differences in ASD (e.g., higher mean pitch and greater pitch variability/range) \cite{fusaroli2017voice,fusaroli2022toward,ma2024can}. For behavior measurement, a neural system quantifying vocal stereotypy achieved session-wise correlations $\geq$0.80 in 6/8 participants (and $\geq$0.90 in several), showing that lapel audio can reliably track repetitive vocal topographies \cite{dufour2020artificial}. In practice, acoustic features (RMS, zero-crossing, spectral centroid/rolloff, MFCCs) augment IMUs when agitation is expressed vocally. Still, agitation often begins without any visible or audible change, prompting a turn to autonomic physiological signals.

Sympathetic arousal (heart-rate increases, Heart-Rate Variability (HRV) reductions, skin-conductance rises) can precede outward behavior by seconds to minutes \cite{critchley2002electrodermal,taelman2009influence,kim2018stress}. These signals are captured noninvasively with Photoplethysmography (PPG)/Electrocardiography (ECG) and EDA following established acquisition/quality practices \cite{allen2007photoplethysmography,park2022photoplethysmogram}. In inpatient/residential contexts, physiology alone has predicted imminent aggression at short horizons: a multi-site study (70 youths, 4 hospitals) reported area under the receiver operating characteristic curve (AUROC) $\approx$0.80 for 3-minute forecasts using logistic regression \cite{imbiriba2023wearable}, while earlier work found AUROC 0.84 (person-specific) vs.\ 0.71 (population) at 1 minute \cite{goodwin2019predicting}. Complementary feasibility data show anticipatory EDA rises in roughly 60\% of agitation episodes \cite{ferguson2019examining}, supporting physiology as a low-salience, autonomic channel, though best used in concert with expressive cues.

How these systems are evaluated shapes what their results mean. A recent systematic review of thirteen studies predicting severe behavior problems from wearables in neurodivergent people finds that methodological concerns reduce the veracity of the advance-prediction claims in this literature, and recommends cross-validation blocked by both participant and time \cite{romani2026using}. Sliding windows are extracted back to back and conventionally overlap by half or more, so adjacent windows share many of the same measurements, and random fold assignment places near-duplicates on both sides of the split \cite{hammerla2015let,dehghani2019quantitative}. Across $47$ clinical wearable-sensor studies, record-wise cross-validation gave a median error of $5.60\%$ against $13.00\%$ for subject-wise \cite{saeb2017need}. The split is therefore a key design choice.

Across these channels, most models have been trained directly on the collected data, whether from a single modality \cite{washington2021activity,cantin2020detecting,dufour2020artificial,imbiriba2023wearable} or several combined \cite{khan2024pilot}, rather than from encoders pretrained outside the task. Foundation models present that opportunity, and concurrent work has begun to take it \cite{kartha2026prediction}. Trained once on a large external corpus, such a model learns general-purpose representations that transfer to downstream tasks with little labeled data, an approach that has reshaped vision, language, and audio and is now reaching wearable and physiological sensing. This fits our setting, where each child contributes only a handful of labeled events, far too few to train an encoder from scratch. Movement encoders pretrained across many human-activity datasets recognize actions they were never trained on \cite{zhang2024unimts}; efficient audio networks pretrained on AudioSet transfer to a broad range of acoustic tasks \cite{schmid2023efficient}; self-supervised models pretrained on large unlabeled photoplethysmography corpora produce embeddings that carry cardiovascular state \cite{pillai2025papagei}; and a model pretrained on a large corpus of wearable electrodermal activity yields embeddings of autonomic arousal \cite{alchieri2026foundation}. Such an encoder can be used with its weights frozen, training only a small head on top, or adapted to the task with low-rank adaptation, which inserts a few trainable parameters into the frozen backbone and tunes it without overfitting the limited data \cite{hu2022lora}. We adopt these models because they are already proven in their respective domains, using them as the encoders our own per-participant data could not train.

Concurrent work applies this transfer to challenging behavior in profound autism, fine-tuning a movement encoder pretrained on a large activity corpus alongside an electrodermal autoencoder and a temperature network, fusing the three, and predicting up to ten minutes ahead in a special education classroom, reaching an AUC of $0.78$ across nine participants \cite{kartha2026prediction}. Earlier work detected the same behaviors from the same three signals without pretraining, reaching an AUC of $0.71$ \cite{rad2025motion}. That line of work predicts the challenging behavior itself, defined as categories shared across a cohort, and we detect the agitation that precedes it, defined by the signs a clinician identified for each child. We also transfer a pretrained encoder for every modality, including audio, which it does not sense. Those numbers are therefore not commensurable with ours, because the label, the negative class, the horizon, and the split all differ.

We bring this transfer to the agitation-detection problem, where it directly addresses the data constraint. The difference from prior work is where the representation comes from. Most systems in this area learn their features from the study's own recordings \cite{khan2024pilot,cantin2020detecting,dufour2020artificial,goodwin2019predicting}, so the representation can be no richer than the labeled events one small cohort provides, and a child who contributes sixty events constrains it as much as they constrain the classifier. We instead take representations learned from external corpora orders of magnitude larger than any agitation dataset, and train a thin projection, fusion block, and per-participant head on top. Each child's few labeled events then specify a decision boundary within a space that is already structured, and no longer have to build that space. The shared structure comes from pretraining, and what remains individual is a linear read-out.

\section{Methods}
\label{sec:methods}
\subsection{Dataset and Data Acquisition}

The data for this study were collected through two sessions of modified IISCA, a structured approach derived from PFA principles. During these sessions, participants were equipped with various wearable technologies to capture relevant data. In the following subsections, we detail the data collection protocol, the sensor modalities utilized, and the characteristics and categorization of the observed signs of agitation. Participant demographics are summarized in Table~\ref{tab:agitation_counts}, with individualized agitation and challenging behavior descriptions provided in Table~\ref{tab:participant_summary} in the Appendix.

\begin{figure}[t]
    \centering
    \begin{subfigure}[b]{0.48\textwidth}
        \centering
        \includegraphics[width=0.8\textwidth]{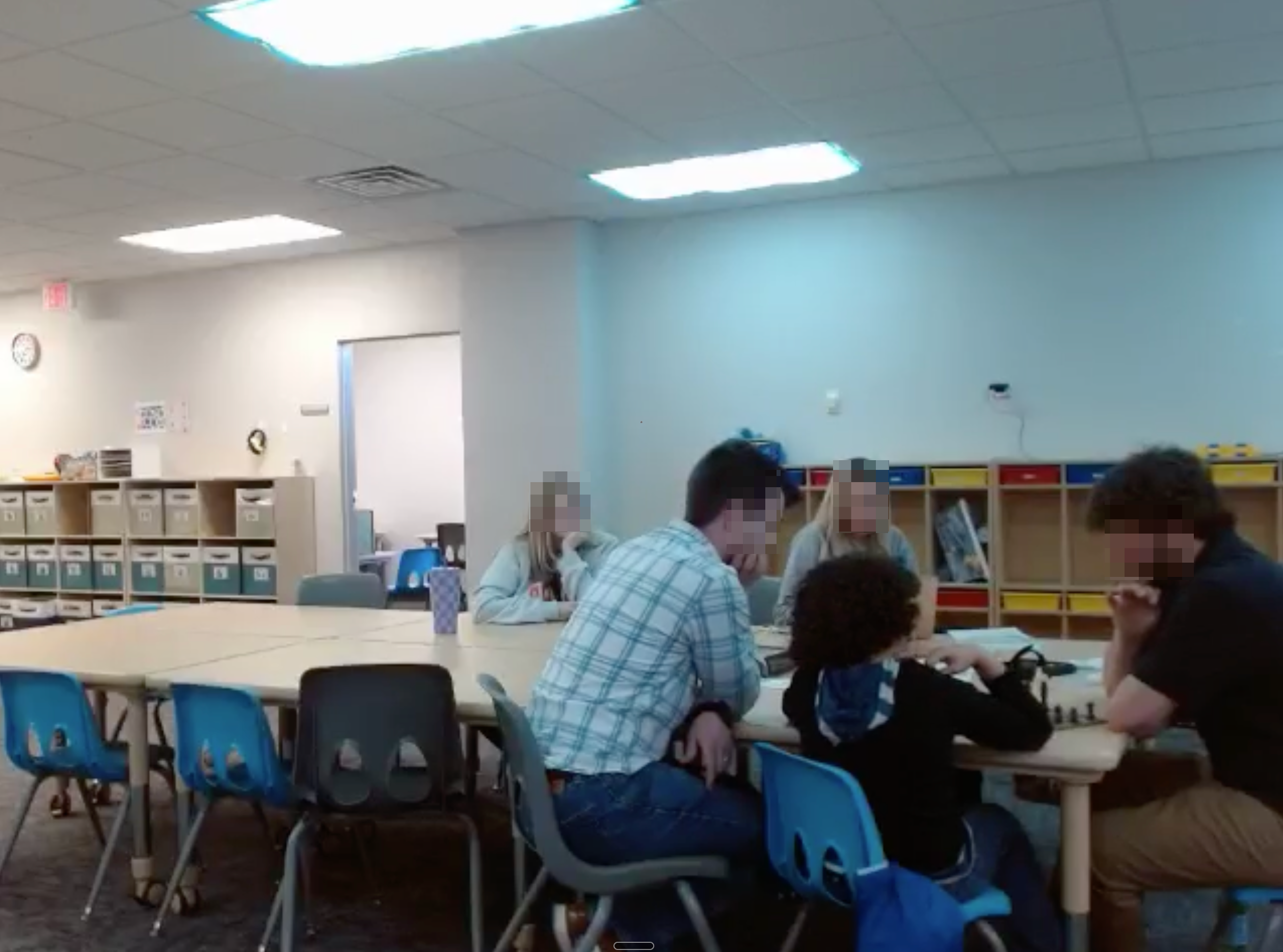}
        \caption{Experimental Room}
        \label{fig:experimental_room}
    \end{subfigure}
    \hfill
    \begin{subfigure}[b]{0.48\textwidth}
        \centering
        \includegraphics[width=0.8\textwidth]{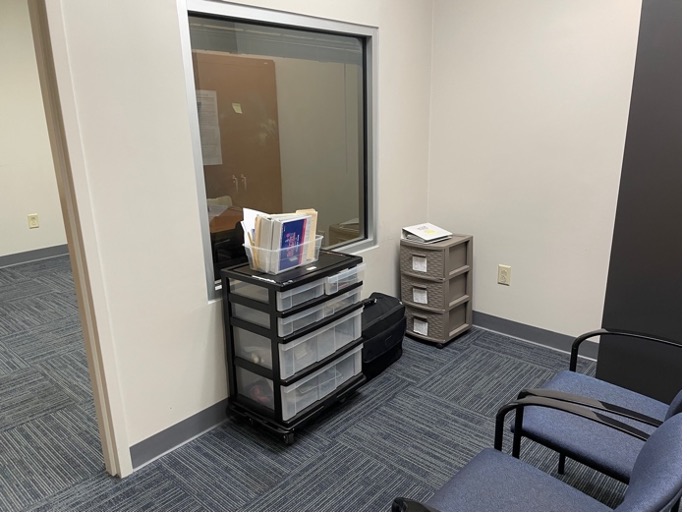}
        \caption{Observation Room \cite{khan2024pilot}}
        \label{fig:observational_room}
    \end{subfigure}
    \caption{Data collection setup, showing two connected spaces: (a) the session room for participants and BCBAs, equipped with sensors, and (b) the observation room for the research team and caregivers, with equipment for monitoring and communication.}
    \label{fig:two-rooms}
\end{figure}

\subsubsection{Data Collection Protocol}

Data collection sessions were conducted in a standardized two-room suite at affiliated therapy centers during normal clinic operations, introducing realistic ambient noise and interruptions (Figure~\ref{fig:two-rooms}). The session room, where the participant and a BCBA engaged in clinician-led sessions, was equipped with highly preferred materials identified through a caregiver interview to establish a ``happy, relaxed, and engaged'' (HRE) state. An adjacent observation room housed the caregiver and a session manager who monitored the session via a one-way mirror or live video feed, confirmed agitation occurrences, and relayed observations to the session-room implementer. Participants and caregivers were informed of their right to pause or terminate at any time; no participant chose to do so. Participants (or caregivers) received \$40 after the first visit and \$90 after the second.

Sessions began with a baseline reinforcement (SR) phase providing unrestricted access to preferred items and social interaction, allowing the participant to acclimate and reach HRE (typically at least 5 minutes). An establishing operation (EO) was then introduced, consisting of individualized antecedent conditions (e.g., removing preferred items, introducing non-preferred demands) designed to evoke non-dangerous agitation. Upon the first clear sign of agitation (or challenging behavior), the EO was terminated and SR reinstated. This contingent alternation enabled repeated, controlled observation of agitation events.

\subsubsection{Ethics and Consent}
All procedures were approved by the Vanderbilt University Institutional Review Board (IRB \#211846). Caregivers provided informed consent, and participants provided assent when able to do so. Sessions were conducted by BCBAs with real-time oversight and a predefined plan to terminate an establishing operation and return to reinforcement conditions immediately upon display of agitation. No adverse events occurred. 

\begin{figure}[t]
    \centering
    \begin{subfigure}[b]{0.32\textwidth}
        \centering
        \includegraphics[width=0.8\textwidth]{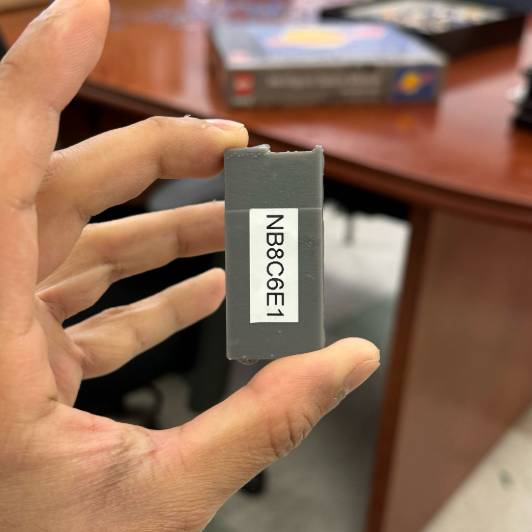}
        \caption{Inertial Measurement Units}
        \label{fig:imus}
    \end{subfigure}
    \hfill
    \begin{subfigure}[b]{0.32\textwidth}
        \centering
        \includegraphics[width=0.8\textwidth]{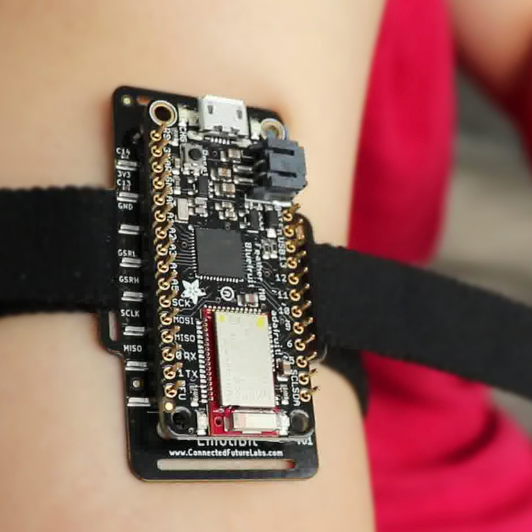}
        \caption{Physiology Sensors \cite{montgomery2023introducing}}
        \label{fig:physiology_sensors}
    \end{subfigure}
    \hfill
    \begin{subfigure}[b]{0.32\textwidth}
        \centering
        \includegraphics[width=0.8\textwidth]{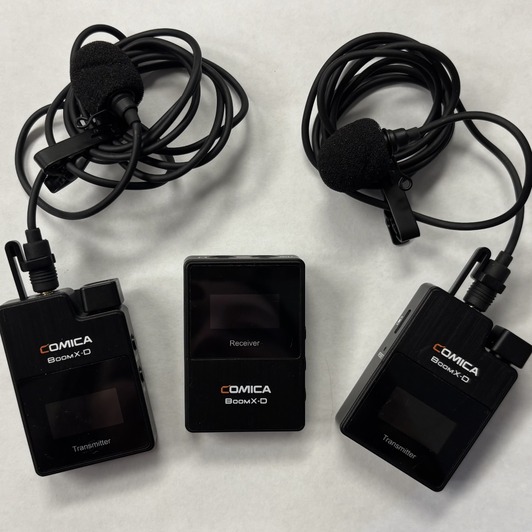}
        \caption{Audio Microphone}
        \label{fig:audio_mic}
    \end{subfigure}
    \caption{Overview of the multimodal sensor array employed for data collection. (a) shows the IMU hardware unit used in the study. (b) displays the wrist- or ankle-worn device used to capture HR, EDA, and acceleration. (c) shows the wireless lavalier microphone used to record vocalizations.}
    \label{fig:sensor_modalities}
\end{figure}

\subsubsection{Modalities}

To capture a comprehensive picture of the participants' states and behaviors during these sessions, a multimodal sensor array was developed and used to record upper-body movement, physiological signals, and vocalizations. The sensor array wirelessly transmitted, through Bluetooth Low Energy (BLE), all data in real-time to a custom-developed iOS application for recording and monitoring.

A key modality, illustrated in Figure \ref{fig:imus}, involved the use of wearable IMUs. Five IMU sensor units were used per participant. These units, with a Bosch BNO055 (9-axis accelerometer, magnetometer, and gyroscope) \cite{sensortec2014intelligent}, were embedded in custom-designed wireless circuit boards. For consistent placement and participant comfort, these circuit boards were inserted into specially designed pockets sewn into a custom shirt worn by the participant. The garment material and attachment decisions were informed by a formative wearability study with autistic youth and caregivers (Section~\ref{subsec:wearability_study}). Placements included one IMU on each wrist, one on each upper arm, and one on the upper torso. The torso IMU was placed either on the upper back or the upper chest depending on participant comfort and tolerance, while keeping the location fixed within a participant across sessions. This configuration was selected to capture both distal and proximal upper-body movement and overall trunk posture, as many individualized signs of agitation and challenging behaviors in our cohort (e.g., hitting, grabbing, slapping surfaces, tensing shoulders, leaning away) are expressed through the arms and upper body, while still keeping all hardware integrated into a single, tolerable garment \cite{min2010automatic, grossekathofer2017automated, morrow2017validation}. Data from all IMUs were sampled at 100 Hz. The accuracy of the IMU data was validated by comparing their derived Euler angles against known positions and orientations.

Physiological data, depicted in Figure~\ref{fig:sensor_modalities}, were captured with a wrist- or ankle-worn EmotiBit, which records electrodermal activity (15\,Hz) and green-channel photoplethysmography (25--50\,Hz), from which heart rate and its variability are derived \cite{montgomery2023introducing}. Not all participants elected to wear the device, in keeping with their sensory profiles, so these signals are available for 7 of the 15 participants, and a modality a participant lacks is masked (Section~\ref{subsec:model}).

Finally, audio data, shown in Figure \ref{fig:audio_mic}, were captured using a wireless lavalier lapel system attached to the participant's clothing to record vocalizations. Audio data were sampled at the typical rate of 48 kHz.

\subsubsection{Formative Wearability Study to Inform Garment and Attachment Design}
\label{subsec:wearability_study}

\begin{figure}[t]
    \centering
    \includegraphics[width=0.8\textwidth, height=0.5\textheight, keepaspectratio]{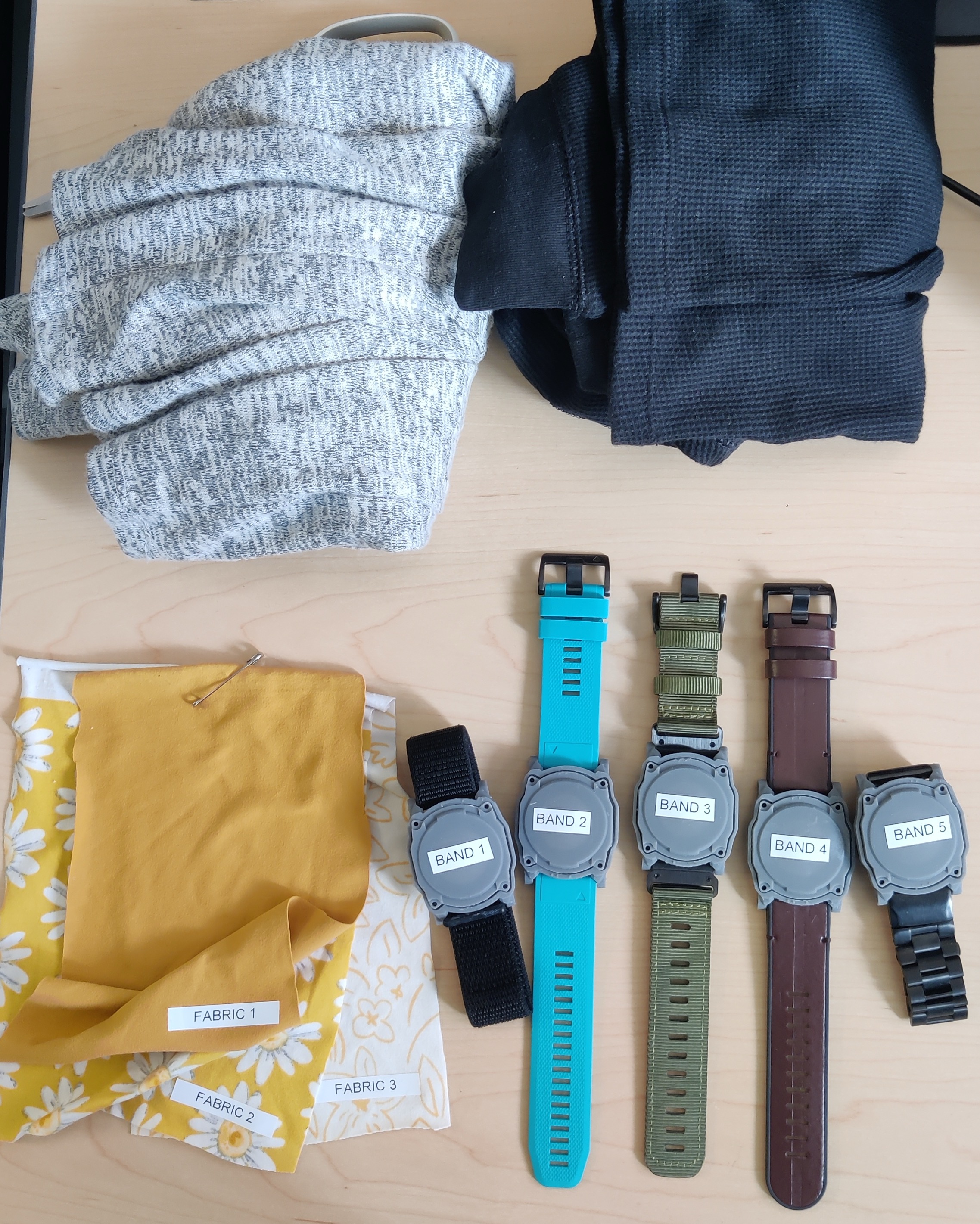}
    \caption{Wearability study components evaluated by participants, including candidate shirt fabrics, attachment bands for a watch-style wearable, and garment prototypes used to assess pocket and sensor-placement preferences.}
    \label{fig:wearability_components}
\end{figure}

Wearability and sensory acceptability are practical determinants of whether autistic youth will tolerate body-worn sensors during real-world use. To inform the garment and attachment mechanisms, we conducted a formative wearability assessment with 7 autistic youth and 6 caregivers. Participants evaluated candidate shirt fabrics, wristband options, and internal vs.\ external sensor pocket designs, providing comfort ratings on a 7-point scale alongside qualitative feedback (Figure~\ref{fig:wearability_components}).

Comfort ratings were highest for the softest fabrics and internal pocket designs, consistent with concerns about scratchiness, seams/tags, and snag risk. Wristband ratings showed greater variability, motivating alternative placements (e.g., ankle) when feasible. These preferences guided the garment material and pocket configuration adopted in our data collection. Although internal pockets were preferred, we used external pockets to enable rapid sensor access during setup and troubleshooting while keeping placement consistent.

\begin{table}[t]
    \centering
    \small
    \begin{tabular}{cccc}
    \hline
    Participant & Age & Sex & Agitation events \\ \hline
    1 & 10 yrs 1 mo & M & 223 \\
    2 & 11 yrs & F & 241 \\
    3 & 11 yrs & M & 366 \\
    4 & 3 yrs 5 mo & F & 81 \\
    5 & 3 yrs 5 mo & M & 66 \\
    6 & 13 yrs 1 mo & M & 333 \\
    7 & 7 yrs 4 mo & M & 211 \\
    8 & 5 yrs 10 mo & F & 81 \\
    9 & 3 yrs 7 mo & M & 80 \\
    10 & 17 yrs 5 mo & M & 64 \\
    11 & 5 yrs 10 mo & F & 106 \\
    12 & 5 yrs 9 mo & M & 61 \\
    13 & 6 yrs 7 mo & M & 58 \\
    14 & 10 yrs 8 mo & M & 196 \\
    15 & 3 yrs 4 mo & M & 96 \\ \hline
    \end{tabular}
    \caption{Participant demographics and number of annotated agitation events. Individualized descriptions are provided in Table~\ref{tab:participant_summary} in the Appendix.}
    \label{tab:agitation_counts}
\end{table}

\subsubsection{Event Annotation and Labeling}

The identification and precise timing of agitation events within the collected data were critical for model training and evaluation. This was achieved through a detailed post-hoc annotation procedure utilizing the synchronized video and audio recordings.

\begin{figure}[t]\centering
    \includegraphics[width=\linewidth]{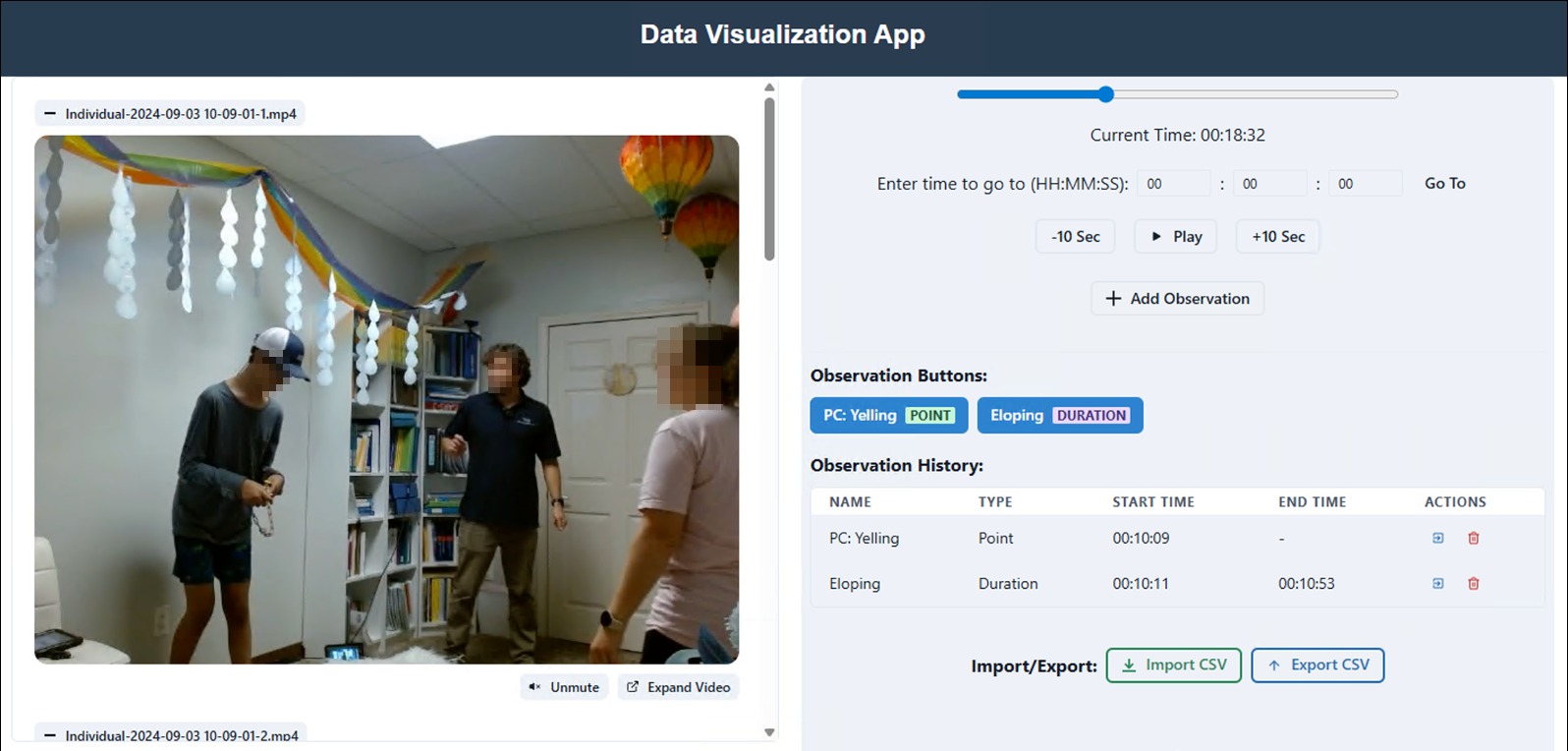}
    \caption{The custom React-based website used to annotate agitation events. Trained annotators reviewed the synchronized four-angle video and lapel audio, navigated frame by frame, and timestamped the onset and offset of each event directly in the browser.}
\label{fig:annotation_site}
\end{figure}

All data collection sessions were recorded using the four webcams positioned in the session room, providing multiple viewing angles, along with the audio from the wireless lapel microphone. These data streams were synchronized with the physiological and motion sensor data. For annotation, these data were loaded into a custom-developed React-based website (Figure~\ref{fig:annotation_site}) \cite{coding_site}. A Registered Behavior Technician (RBT) or a BCBA used the website to review the session recordings and label the occurrences of agitation. Each identified event was timestamped directly within the website. The platform facilitated frame-by-frame video navigation and audio playback control to enhance the accuracy of temporal annotations. Inter-observer agreement (IOA) was calculated for 20\% of all recorded video sessions. This involved a second independent trained annotator coding these selected sessions. Agreement for agitation event occurrence (presence/absence in a defined window) and onset/offset timing (within a ±1-second tolerance) was assessed using Cohen’s Kappa, with values consistently exceeding $0.85$. Disagreements were resolved through discussion and consensus with a supervising BCBA.

The frequency of annotated agitation varied widely across the cohort, reflecting the idiosyncratic nature of behavioral escalation. Table~\ref{tab:agitation_counts} reports the number of annotated agitation events per participant, highlighting the inter-individual variability in how signs of agitation manifest.

\subsubsection{Pooling Individualized Signs}

As introduced in Section~\ref{sec:introduction}, we treat agitation as a state defined by the presence of individualized, clinically identified signs of distress, and our task is to detect those signs. Agitation is expressed differently by every participant, and the individualized signs are presented in Table~\ref{tab:participant_summary}. Treating each sign as its own class would spread limited data across many labels and yield unstable performance, particularly for the rarer signs. We therefore pool every sign into a single agitation class against a non-agitation background, and the model detects that class rather than any individual sign.

We next describe how these raw, continuous streams become the features and inputs the model uses.

\subsection{Signal Processing and Feature Engineering}

Continuous data streams from all modalities were segmented into fixed windows, from which we extracted the engineered features described below along with the raw signals consumed by the foundation models (Section~\ref{subsec:model}). Because our data-collection protocol's contingent alternation repeatedly evokes agitation and ends the evoking condition at the first sign, agitation onsets recur within a session, so each window must be short enough to isolate one. A longer window would enclose more than one, breaking the correspondence between a window and a single agitation event. The same localization keeps the lead-time analysis meaningful (Section~\ref{subsec:advance_prediction}), where the label is shifted to windows before the onset and a longer window would blur advance detection together with detection at the onset. We set the length to $15$\,s because an agitation episode plays out over several seconds of movement and vocalization, long enough to capture that dynamic and absorb the annotation-timing jitter while keeping the onset localized without absorbing prior windows of agitation. 

\subsubsection{Feature Engineering}

Each modality gives the model two inputs: the raw signal, which a pretrained foundation model encodes into an embedding (Section~\ref{subsec:model}), and a set of hand-crafted features computed within the window. All four modalities use both, concatenated before projection. The engineered features are compact, interpretable descriptors computed directly on the window. For movement, photoplethysmography, and electrodermal activity, they carry a signal the foundation-model embedding does not otherwise receive, and for audio they summarize the lapel-microphone signal with standard spectral and energy descriptors.

For movement, the engineered features summarize upper-body IMU motion. They include per-location speed statistics (mean, standard deviation, maximum, 90th percentile, and energy) and shape descriptors that separate discrete actions from ambient movement (jerk, kurtosis, crest factor, peak rate, and spectral entropy). We also add per-axis angular velocity and energy \cite{saraf2023survey,ahmad2022survey}. For electrodermal activity, we compute arousal features from skin conductance: mean, variability, range, and slope. For photoplethysmography, we compute cardiac features: mean and variability of heart rate, heart-rate-variability metrics SDNN, RMSSD, and pNN50, and pulse amplitude. For audio, we compute standard descriptors of the lapel-microphone signal: root-mean-square energy, zero-crossing rate, spectral centroid, rolloff, and bandwidth, and the first thirteen mel-frequency cepstral coefficients.

\subsubsection{Feature Set Construction}

The engineered features were aggregated into a per-modality feature vector. Missing engineered values were imputed with zero prior to z-score normalization. We standardized each engineered feature per participant using that participant’s mean ($\mu$) and standard deviation ($\sigma$): $z=(x-\mu)/\sigma$. The foundation-model embeddings are used as produced, without per-participant standardization. Normalization is instead applied inside the model, by the layer normalization in each modality’s projection (Section~\ref{subsec:model}). Note that an imputed zero in raw space becomes $(0-\mu)/\sigma$ after standardization (not necessarily $0$). This assigns missing entries a constant, data-driven value without fabricating temporal structure. We prefer this over forward-fill because forward-filling windowed statistics (e.g., variance, peak counts, short-term trend) can introduce artificial persistence and autocorrelation, precisely where rapid changes convey signal near agitation onsets. Zero-impute with z-score preserves marginal distributions while avoiding synthetic trajectories. To prevent leakage, $\mu$ and $\sigma$ are computed from training data only (per participant) and reused for validation/test.

\subsubsection{Target Label Generation for Predictive Modeling}
\label{subsec:target_labels}

For each participant we build a labeled set of $15$\,s windows. Each clinician-annotated onset contributes one agitation window, the $15$\,s window ending at that onset. Non-agitation windows are sampled on a $15$\,s grid from stretches at least $15$\,s from any onset. We report a single detection task over all 15 participants, signs of agitation versus non-agitation. Engineered features are z-scored per participant from training-fold statistics, so each feature reflects deviation from that child's own baseline, and $\mu$ and $\sigma$ are never computed from held-out windows.

An annotation marks the moment a clinician could first identify and timestamp a sign of agitation from video. It does not mark the moment agitation began. Agitation builds over seconds and includes autonomic components that no observer can see, so the annotated onset is best understood as an upper bound on true onset: agitation is already underway when it is marked. We treat this as an explicit modeling assumption throughout, and it is the same assumption used to set label window length in prior work \cite{khan2024pilot}. It has a direct consequence for evaluation. If agitation is present before it is annotated, then a model that flags a window shortly before the annotation may be detecting the state rather than making an error, and standard onset-aligned metrics will score those detections as false positives.

To study how performance changes as a function of lead time, we repeated the evaluation with the input window positioned earlier relative to each onset. For a lead of $\Delta t$, the $15$\,s window ends $\Delta t$ seconds before the annotated onset ($\Delta t = 0$ ends at the onset), with its label unchanged; we evaluate $\Delta t \in \{5, 10, 15, 20, 25, 30\}$\,s. This measures how well the model separates pre-onset windows from non-agitation windows. They do not establish that the state present at $\Delta t$ was independently verified as agitation by a clinician, because no such annotation exists for those windows. Section~\ref{subsec:advance_prediction} reports these results, and Section~\ref{sec:discussion} discusses the interpretation this supports.

\subsection{Multimodal Foundation-Model Architecture}
\label{subsec:model}

\begin{figure}[t]
    \centering
    \includegraphics[width=\textwidth]{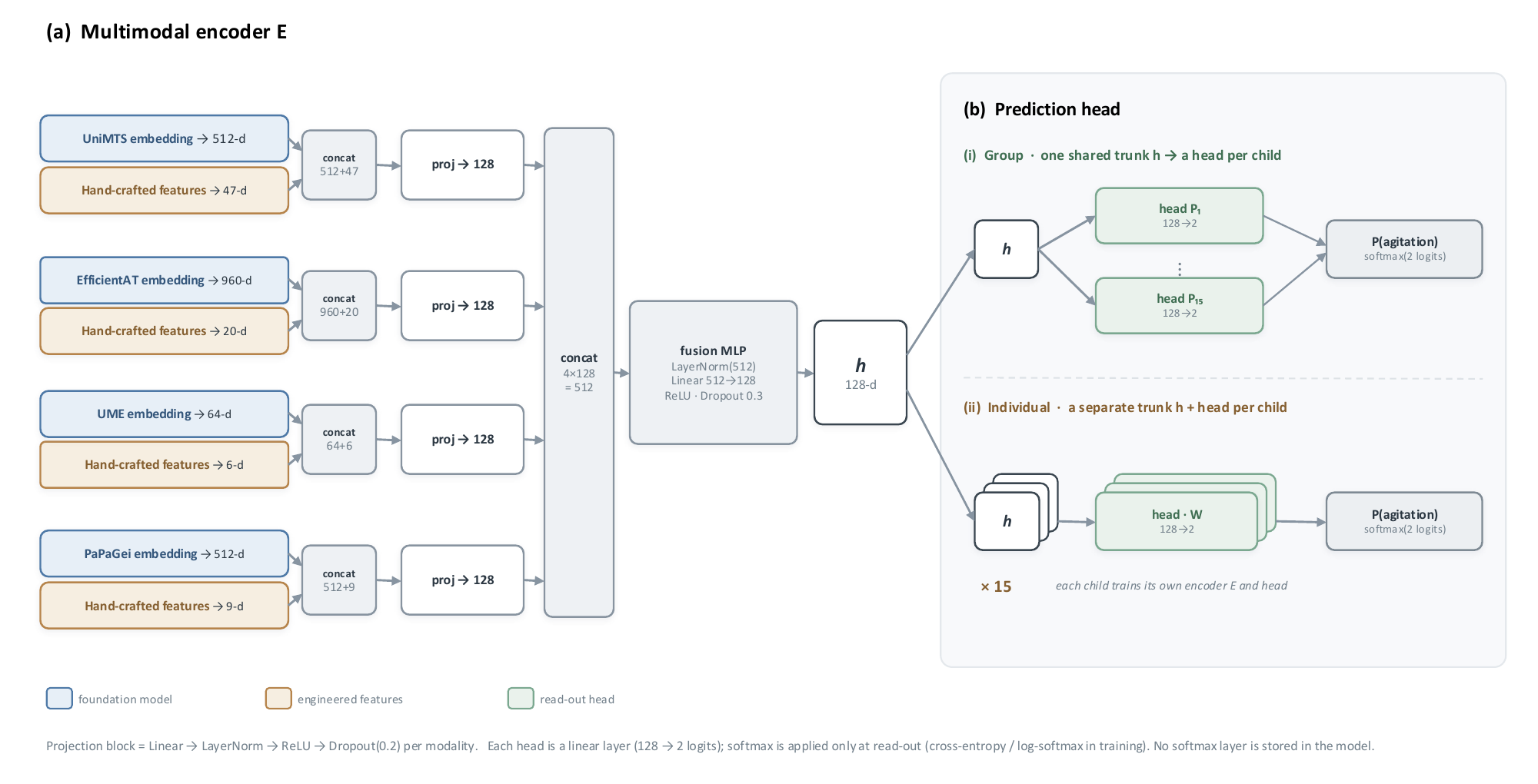}
    \caption{Multimodal foundation-model architecture. \textbf{(a)}~Each 15\,s modality window is encoded by a pretrained foundation model (UniMTS for movement, EfficientAT for audio, PaPaGei for PPG, and UME for electrodermal activity), concatenated with that modality's engineered features, and projected to a shared 128-dimensional space. The four projections are fused into a shared representation~$h$. \textbf{(b)}~The two model forms differ only at the read-out head. The \emph{group} model shares one trunk and uses a separate linear head per participant, selected by participant identity. The \emph{individual} model trains a separate trunk and head per participant, sharing nothing. Each head outputs two logits, and softmax gives the agitation probability at read-out and no softmax layer is stored.}
    \label{fig:architecture}
\end{figure}

Each participant contributes few labeled signs of agitation, so rather than learn a representation from scratch we transfer four foundation models pretrained on large external corpora, one per modality. Movement is encoded by a spatio-temporal graph network pretrained across many human-activity datasets \cite{zhang2024unimts}, photoplethysmography by a self-supervised model \cite{pillai2025papagei}, audio by an efficient convolutional network pretrained on AudioSet \cite{schmid2023efficient}, and electrodermal activity by a convolutional network pretrained on a large corpus of wearable electrodermal recordings \cite{alchieri2026foundation}.

\subsubsection{Per-modality projection and fusion}
Each foundation model produces a modality embedding of a different size (512, 512, 960, and 64 dimensions for movement, photoplethysmography, audio, and electrodermal activity, respectively). Each modality embedding is concatenated with that modality's engineered features. Each modality's representation is projected to a common 128-dimensional space through a linear layer, layer normalization, a rectified linear unit, and dropout. Projecting every modality to the same width keeps any one embedding from dominating the fusion. The four projected modalities are concatenated and passed through a fusion block and a classification head that outputs an agitation probability (Figure~\ref{fig:architecture}a). We use a group formulation in which a single trunk is shared across all participants and a per-participant head reads out the fused representation, so that shared structure is learned jointly while each child retains an individualized decision boundary (Figure~\ref{fig:architecture}b). Modalities a given participant lacks (for example physiology for a child who wore no device) are masked to zero, and the per-participant head learns to weight the modalities that participant actually has.

\subsubsection{Transfer configurations}
We evaluate three ways of using the pretrained backbones. In the \emph{frozen} configuration the foundation models are held fixed and their embeddings are precomputed once, and only the projections, fusion block, and head are trained. In the \emph{fine-tuned} configuration the pretrained weights are adapted with low-rank adaptation \cite{hu2022lora}: the backbone weights stay frozen and small low-rank adapters are inserted into the convolutional and linear layers, so that only a few hundred thousand parameters are trained rather than the full backbones. In the \emph{from-scratch} configuration, the backbones keep the same architecture but are randomly initialized and trained end to end, isolating the contribution of pretraining. We evaluate each configuration in both the group form above and an \emph{individual} form in which a separate model is trained for each participant, giving a two-by-three comparison across $\{$group, individual$\}$ and $\{$frozen, from-scratch, fine-tuned$\}$.

\subsection{Experimental Design and Evaluation Protocol}

We evaluate every configuration the same way, so the comparisons that follow differ in the model and not in how it was scored. This section covers the cross-validation protocol, the training setup, the metrics, and the specific configurations we compare.

\subsubsection{Training and Testing Strategy}
\label{subsubsec:cv}

Time-series physiological and kinematic data are highly autocorrelated: windows close in time share signal, so randomly assigning windows to training and testing would place near-duplicate windows on both sides of the split and leak information across it \cite{hammerla2015let,saeb2017need}. We therefore evaluate each participant with five-fold chronological block cross-validation. We order each child's session timeline in time and partition it into five contiguous blocks. Each fold holds out one block for testing and trains on the other four, an $80/20$ split, and every block is held out exactly once. We use five folds rather than a single $80/20$ split because a participant's signs of agitation are heterogeneous and spread unevenly across the session. Any one held-out block may contain some kinds of signs but not others. Rotating the held-out block through all five folds tests every window. Per-participant standardization statistics are computed from the training folds only, and all reported metrics come from these held-out predictions. We adopt this protocol rather than the random, class-balanced split behind the highest accuracies reported in this literature, which a recent review of wearable prediction studies in neurodivergent populations recommends against for this reason \cite{romani2026using}. That split reports a much higher AUC on our own data, and we give the comparison in Section~\ref{subsec:split_protocol}.

\subsubsection{Model Training Details}

All models were trained with the Adam optimizer and a class-weighted cross-entropy loss to account for the imbalance between agitation and non-agitation windows. In the fine-tuned configuration the low-rank adapters were trained at a learning rate of $1\times10^{-3}$; in the from-scratch configuration the backbones were trained at $1\times10^{-4}$; the projections, fusion block, and heads were trained at $1\times10^{-3}$ throughout. Training used mixed-precision arithmetic. Because the fine-tuned and frozen configurations start from pretrained weights, they converge in a small number of epochs rather than the long schedules required to train comparable models from random initialization.

\subsubsection{Performance Metrics}

We report the area under the receiver operating characteristic curve (AUC) as the primary metric, because it is threshold-independent and directly measures how well agitation windows are ranked above non-agitation windows. For each participant we compute AUC over that child's held-out windows, and summarize across the cohort as the macro average, weighting every participant equally. We attach a 95\% bootstrap confidence interval to every reported AUC, resampling windows within a participant for per-participant intervals and resampling participants for the cohort mean. To test whether the cohort mean exceeds chance, we use a within-participant label-permutation null: within each participant, agitation and non-agitation labels are shuffled, the full pipeline is re-run, and the observed macro AUC is compared against the resulting null distribution.

\subsubsection{Experimental Configurations}

Our experiments answer four questions about detecting agitation in this cohort. First, how well does the model detect agitation at the clinician-annotated onset, and does detection require the external pretraining or a separate model per child? Second, how far in advance can it detect agitation, evaluated by shifting labels earlier in 5\,s steps up to 30\,s? Third, which sensing modalities carry the signal, evaluated by dropping each in turn? Fourth, is a watch-only configuration sufficient, evaluated on the subset whose watch recorded those signals?

\section{Results}
\label{sec:results}

We report all results over the 15 participants as a whole. Unless stated otherwise, numbers are macro AUC across participants for the group, fine-tuned model, with 95\% bootstrap confidence intervals.

\subsection{Detection at the Annotated Onset}
\label{subsec:onset}
At the clinician-annotated onset, the group fine-tuned model reached an AUC of $0.724$ (95\% CI $0.650$--$0.794$; pooled across all windows, $0.785$). The median across participants is $0.710$, robust to the two participants whose models did not clear chance. A within-participant permutation test placed this well above chance. Thirteen of fifteen participants were individually above chance. Of the remaining two, one participant's confidence interval crosses $0.5$ and the other falls entirely below it (Figure~\ref{fig:forest}; per-participant values in Table~\ref{tab:participant_results}).

\subsection{Pretraining and the Group Formulation}
\label{subsec:transfer}
\begin{figure}[t]
\centering
    \includegraphics[width=0.6\linewidth]{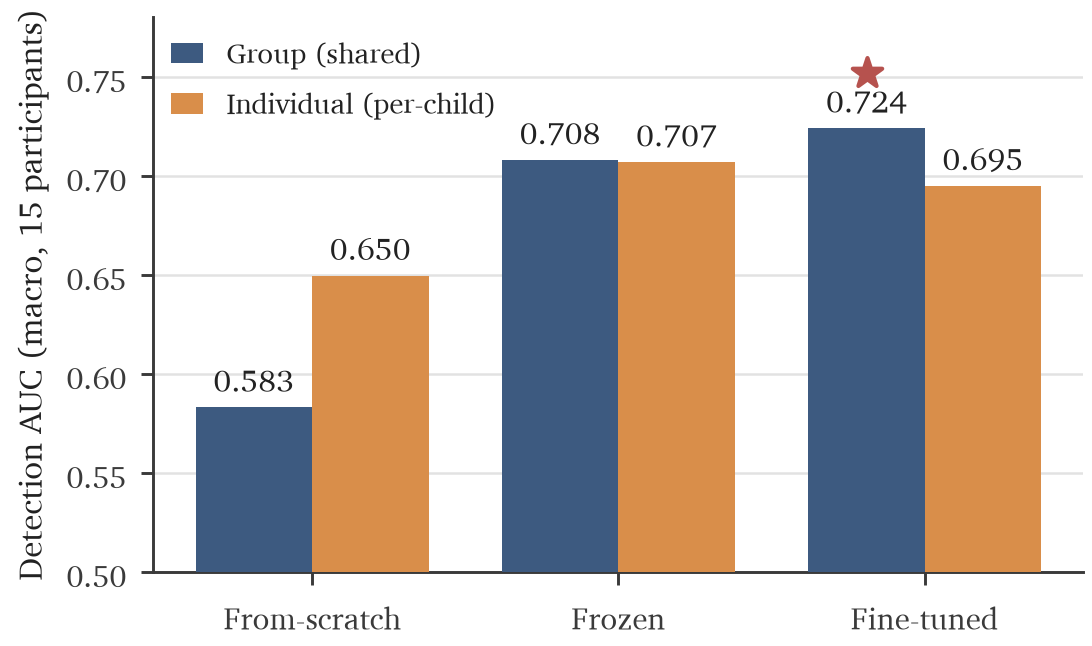}
    \caption{Detection AUC across all 15 participants for the two-by-three transfer comparison. Pretraining (frozen, fine-tuned) is far above from-scratch. The group model is best overall and matches per-participant models closely.}
\label{fig:grid}
\end{figure}

Table~\ref{tab:grid} and Figure~\ref{fig:grid} report the two-by-three comparison. Pretraining was valuable as the from-scratch group model reached only $0.583$, while both frozen ($0.708$) and fine-tuned ($0.724$) foundation-model results were far higher. Fine-tuning with low-rank adapters gave a gain over frozen features that fell within the confidence interval. In the individual family the ordering even reversed (frozen $0.707$ versus fine-tuned $0.695$). The pretrained representations are therefore already close to their ceiling on this task, and adaptation adds little. A single group model was best overall ($0.724$), and its confidence interval overlaps the individual model's in every configuration (Table~\ref{tab:grid}), so the group-versus-individual differences are within sampling noise. Per-child training is therefore unnecessary for detection.

\subsection{Detection Before the Annotated Onset}
\label{subsec:advance_prediction}
Figure~\ref{fig:forecast} shows detection AUC as the labels are shifted earlier, for the best model of each family (group fine-tuned and individual frozen). Performance is strongest at onset and decays smoothly. The group model declines from $0.724$ at $0$\,s to $0.632$ at $15$\,s and $0.608$ at $30$\,s; the individual model declines from $0.707$ to $0.630$ to $0.586$. The group model is the stronger detector at the annotated onset and stays at or above the per-child individual model across the forecasting horizon, with the two nearly coinciding through the mid leads. Both stay above the $0.5$ chance level throughout. We tested significance at the annotated onset and report the earlier leads as point estimates. As in prior work, we cannot fully separate detection of an earlier state from temporal autocorrelation with the annotated window.

\subsection{Which Modalities Carry the Signal}
\label{subsec:ablation}
\begin{figure}[t]
    \centering
    \includegraphics[width=0.6\linewidth]{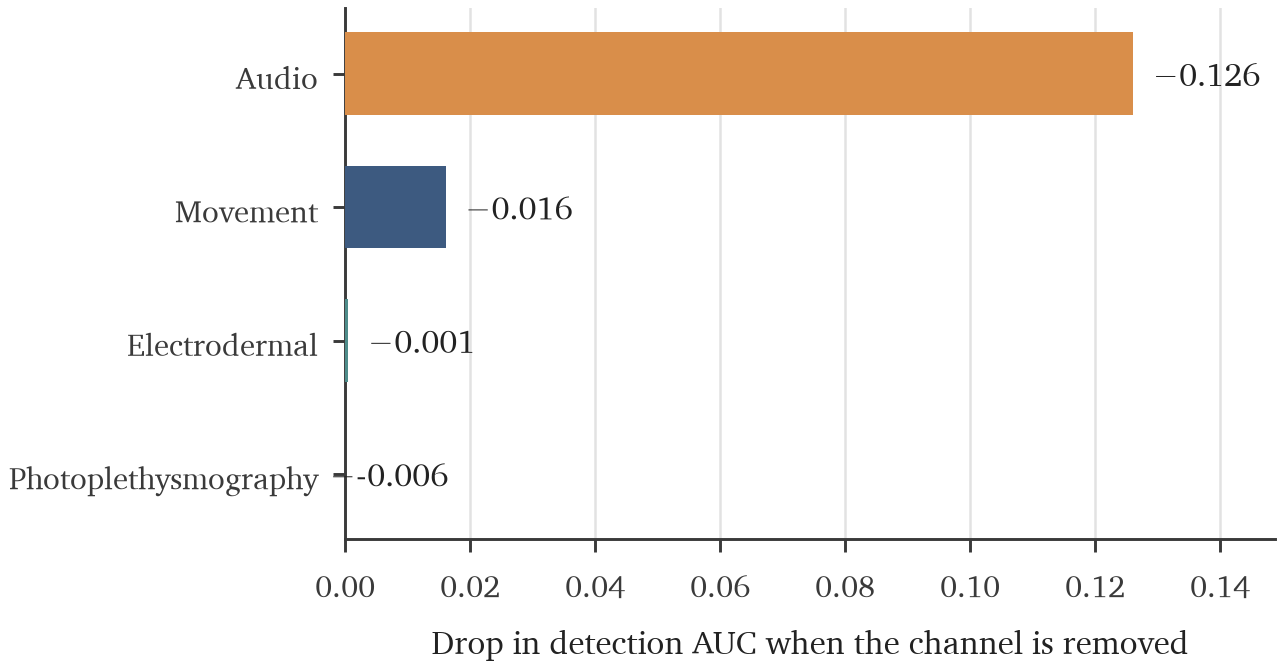}
    \caption{Drop in detection AUC when each modality is removed. Audio accounts for most of the signal.}
    \label{fig:ablation}
\end{figure}

Dropping each modality in turn (Figure~\ref{fig:ablation}) shows that audio dominates. Removing it cost $0.126$ AUC, while removing movement, electrodermal activity, or PPG each changed it by less than $0.02$. Audio, which captures vocal expression through the lapel microphone, is the primary modality through which the model recovers agitation in this cohort. This is a cohort average, and it hides wide variation across children: the cost of removing audio ranges from $0.036$ to $0.351$ per participant, and the near-zero aggregate cost of the other channels averages over children who rely on them and children who do not.

\subsection{Watch-Only Feasibility}
\label{subsec:watch}
\begin{figure}[t]
    \centering
    \includegraphics[width=0.6\linewidth]{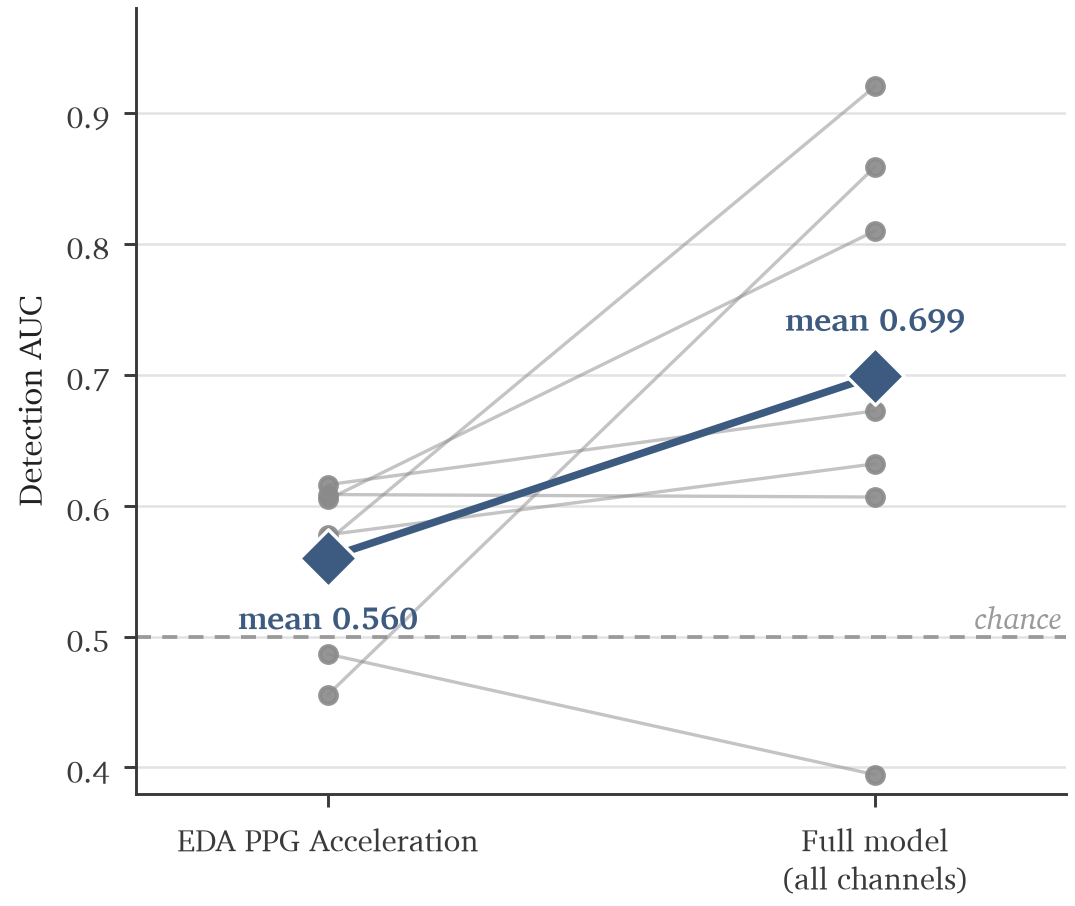}
    \caption{Watch-only detection on the 7 participants whose watch recorded electrodermal activity, photoplethysmography, and acceleration. The watch-only configuration stays near chance ($0.560$) but well below the full model ($0.699$).}
\label{fig:watch}
\end{figure}

On the 7 participants whose watch recorded electrodermal activity, photoplethysmography, and acceleration, a watch-only configuration reached only $0.560$, far below the full multimodal model's $0.699$ on the same participants (Figure~\ref{fig:watch}). Two of the 7 fell below chance. A low-profile wearable therefore provides at best a weak signal for this task, and the lapel microphone remains necessary for strong detection of individualized agitation.

\subsection{Comparison with Baseline Methods}
\label{subsec:baselines}
The from-scratch configuration serves as a deep baseline that shares the model's architecture but forgoes pretraining, and it reached only $0.583$, below both frozen and fine-tuned transfer. Pretrained representations, and the audio modality in particular, account for most of the detectable signal.

\subsection{Effect of the Evaluation Split}
\label{subsec:split_protocol}
All results reported above come from five-fold chronological block cross-validation. Under the protocol behind the highest performance in this literature, treating each window as an independent observation with balanced classes and a random split, the same model reaches an AUC of $0.994$ and $97.0\%$ accuracy on the same data. This difference depends the split strategy as adjacent windows share signal and a random assignment places near-duplicates of the held-out windows into the training set, scoring the model on windows it has effectively already seen (Section~\ref{subsubsec:cv}). We report the chronological result because it estimates performance on a session the model has not seen, and results obtained under record-wise splits are therefore not directly comparable to ours.

\begin{table}[t]
    \centering
    \small
    \setlength{\tabcolsep}{5pt}
    \begin{tabular}{lccc}
    \hline
     & From-scratch & Frozen & Fine-tuned \\ \hline
    Group      & 0.583 [.53,.64] & 0.708 [.64,.77] & \textbf{0.724 [.65,.79]} \\
    Individual & 0.650 [.59,.71] & 0.707 [.64,.77] & 0.695 [.63,.76] \\ \hline
    \end{tabular}
    \caption{Detection AUC at the annotated onset across all 15 participants (macro average), with 95\% participant-bootstrap confidence intervals, for the two-by-three comparison of $\{$group, individual$\}$ $\times$ $\{$from-scratch, frozen, fine-tuned$\}$. Group and individual intervals overlap within each column, and the best cell (group fine-tuned) has a within-participant permutation $p=0.0005$. Per-participant confidence intervals (Figure~\ref{fig:forest}) are reported for the group model only.}
    \label{tab:grid}
\end{table}

\begin{figure}[t]
\centering
    \includegraphics[width=0.6\linewidth]{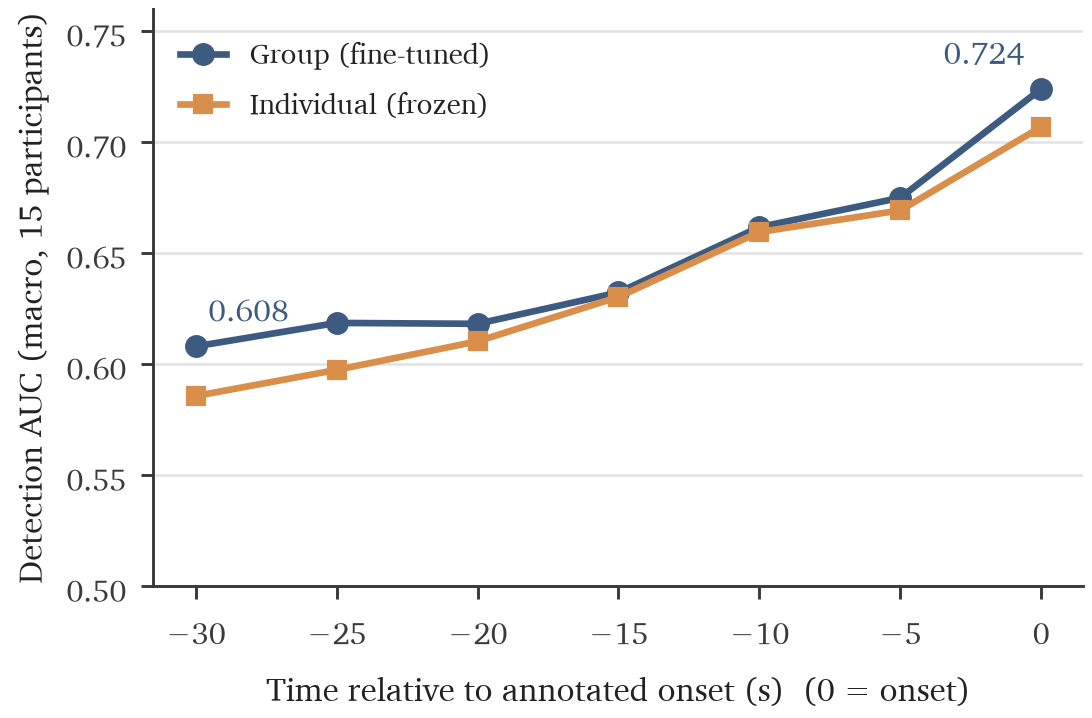}
    \caption{Detection AUC as labels are shifted earlier than the annotated onset, for the best model of each family. Both decline steadily but stay above chance through $30$\,s. The group model leads at the annotated onset and remains at or above the individual model across the horizon.}
\label{fig:forecast}
\end{figure}

\begin{figure}[t]
\centering
    \includegraphics[width=0.6\linewidth]{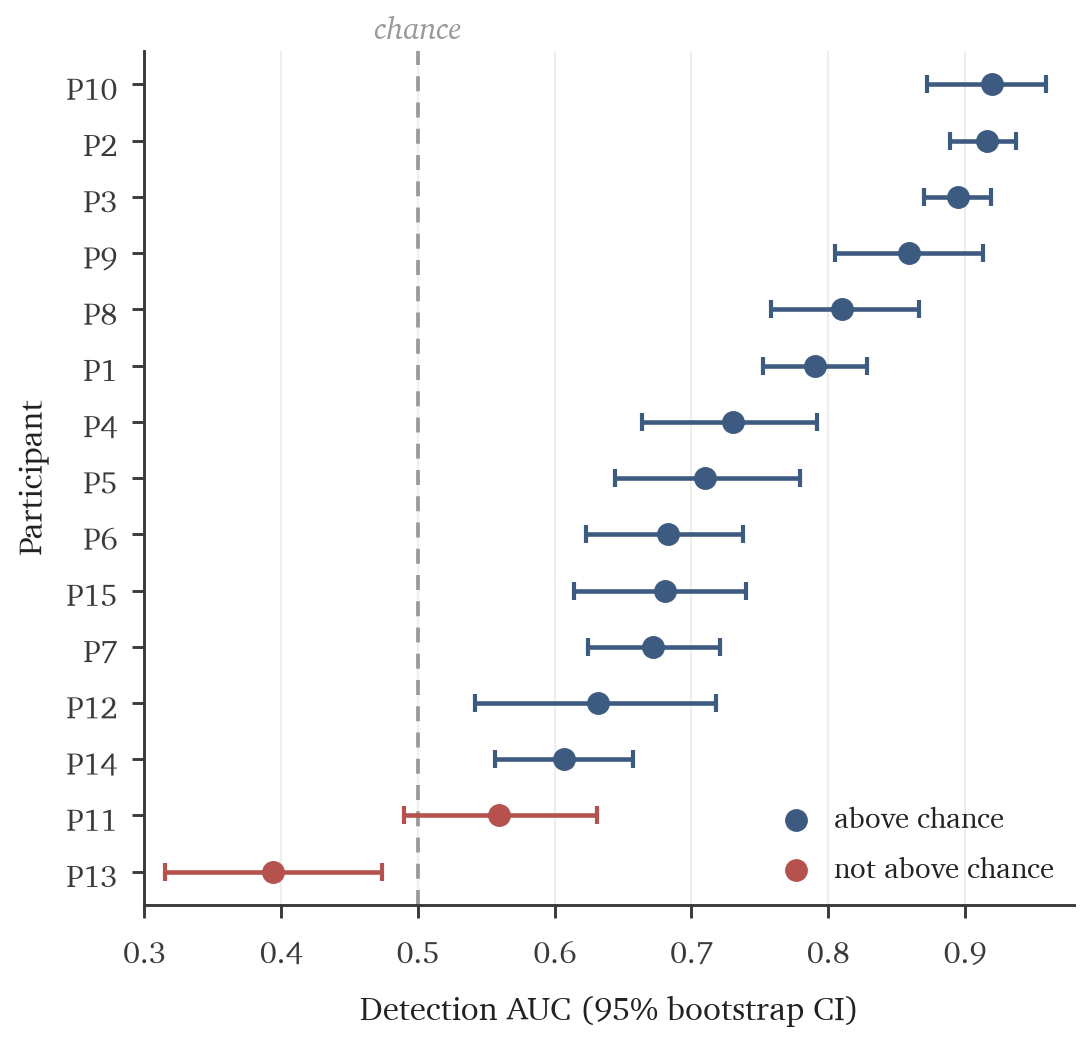}
    \caption{Per-participant detection AUC with 95\% bootstrap confidence intervals (group fine-tuned). Thirteen of fifteen participants are above chance. One participant's interval crosses $0.5$ and one falls entirely below it.}
    \label{fig:forest}
\end{figure}

\section{Discussion}
\label{sec:discussion}

We adapted four pretrained foundation models into a single shared model and evaluated how well it detects individualized agitation and how far in advance, along with how much each sensing modality and the pretraining itself contribute.

\subsection{Interpreting Model Behavior and Individualized Prediction}

The model's behavior raises two questions worth unpacking: why detection quality varies so widely across children, and what the pretraining and parameter sharing actually contribute.

\subsubsection{Heterogeneity Across Participants}

Detection quality varied widely across participants (Figure~\ref{fig:forest}), from AUCs above $0.9$ for two children down to one whose detection fell below chance. Some of this variability tracks how each child expresses agitation (Table~\ref{tab:participant_summary}). Children whose signs are clear and discrete tend to separate well from non-agitation windows, while detection is weaker for children whose signs are more subtle. The pattern has exceptions, and some children with overt signs still score poorly.

For this analysis we label each annotated window by whether the agitation was primarily vocal, a discrete movement, or a diffuse ``other'' sign that is neither (e.g., subtle postural or affective changes, withdrawal). At the window level the split is clear: discrete movement and vocal signs are detected well (AUC $0.813$ and $0.786$), while diffuse ``other'' signs are much harder ($0.731$). This is the audio-dominance result from the other direction, since vocal signs land squarely in the microphone, the modality that carries most of the signal, while the diffuse signs that no channel captures cleanly are where the model struggles. Between children the same tendency does not always hold. A child's number of diffuse signs correlates only loosely with their AUC (Spearman $\rho = -0.40$, not significant at fifteen children). For some participants, we get unexpected results: Participant 13, the one child below chance, showed loud grunts, crying, and forceful kicking and throwing rather than quiet signs, and Participant 14, among the most vocal children in the cohort, still scored well below the participants the model detected most reliably. For them the limit is not that their signs are subtle but more likely the quality of the negative class (both had the lower-arousal proxy baseline discussed below) and how well the sensors fit their particular behaviors.

The sensor ablation shows the same divide per child. Removing audio costs a group-level $0.126$ in AUC, but the per-participant loss ranges from $0.036$ to $0.351$. Nine of the fifteen lose more than $0.10$ without audio, while for the rest movement and physiology carry the signal. The multimodal stack is therefore not redundant across the cohort.

A separate factor is the quality of the negative class. The eight children with a clinician-verified calm baseline averaged $0.776$, against $0.665$ for the seven whose negative windows came from a lower-arousal proxy (Mann--Whitney $p = 0.15$); part of the difficulty is label noise, not the child. What does not explain performance is who the child is or how much data they gave: detection AUC is uncorrelated with the number of annotated windows per participant (Spearman $\rho = 0.23$), with age ($\rho = 0.31$), and with sex ($0.754$ AUC for the four girls versus $0.713$ for the eleven boys). None of these participant-level relationships, agitation sign type and baseline included, individually clears significance at fifteen children, so we treat them as a set of trends rather than established effects.

A shared model suits most of these children, but not all. The group model beats its per-child counterpart for 11 of the 15 participants, by a small margin on average ($0.02$ AUC). The exceptions are children whose own build-up appears more informative than cohort-level structure.

\subsubsection{Score Separation and Failure Modes}
\begin{figure}[t]
    \centering
    \includegraphics[width=\linewidth]{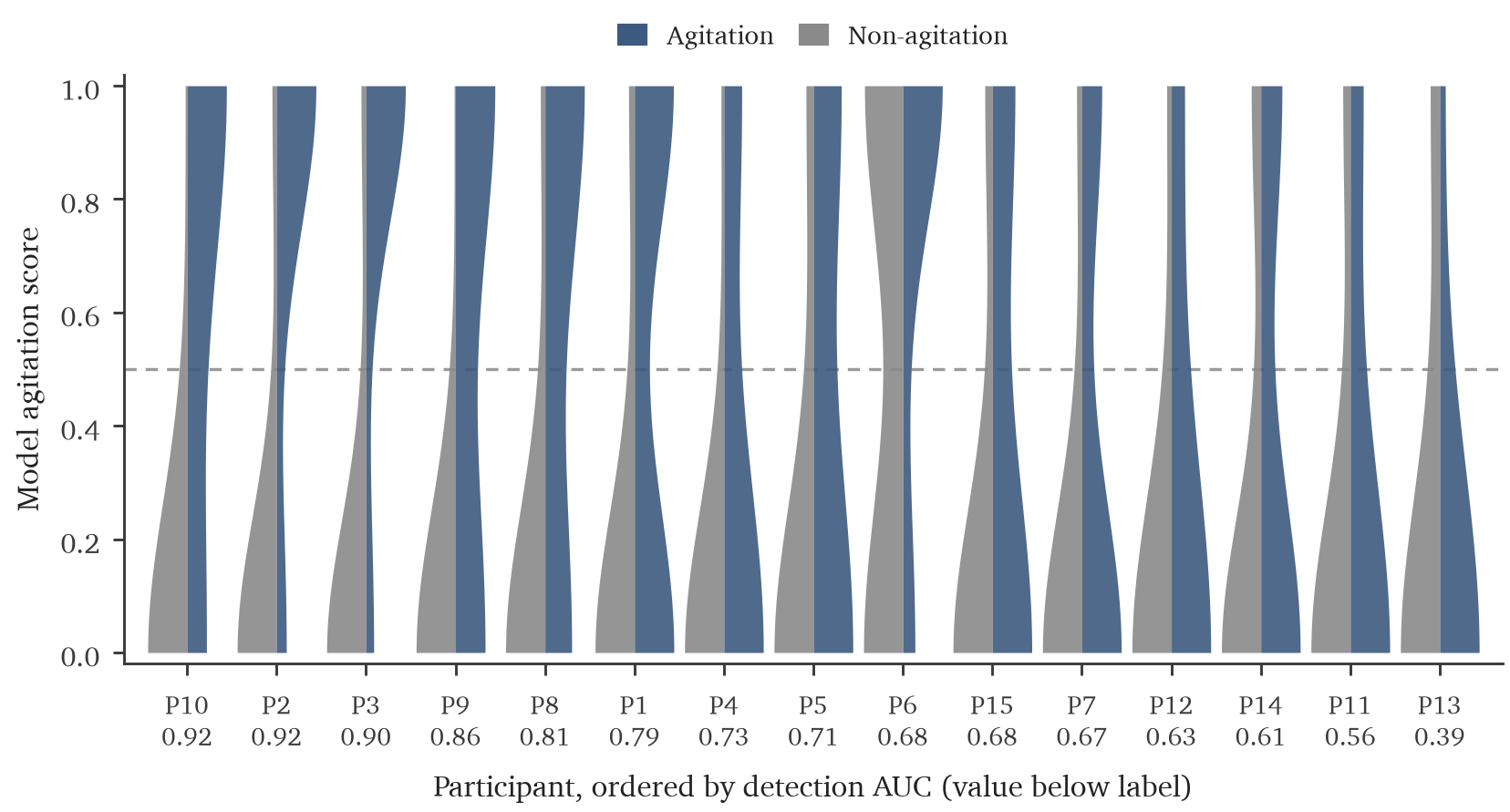}
    \caption{Per-participant separation of the model's output scores at the annotated onset, agitation (blue) versus non-agitation (gray), ordered by detection AUC. Well-detected children show cleanly split distributions, and those it detects poorly overlap. Two failure modes are visible. Participant 6 pushes many non-agitation windows to high scores (false alarms), while for Participant 13 the agitation and non-agitation scores overlap almost entirely and its detection falls below chance.}
\label{fig:score_sep}
\end{figure}

Figure~\ref{fig:score_sep} shows the model's output score for every window, split by the true state and grouped by participant. Each participant is one split violin. Its right half (blue) is the model's score distribution on that child's agitation windows, its left half (gray) on the non-agitation windows. The vertical axis is the score from $0$ to $1$, with the dashed line at $0.5$. Participants are ordered left to right by decreasing detection AUC, the same ranking as in Figure~\ref{fig:forest}, with each child's AUC printed below the label. A child is separated cleanly when the blue half piles up near the top and the gray half near the bottom. Where detection is weakest, the two halves overlap across the range. The plot also separates two failure modes that a single AUC hides. For some children, such as Participant 6, the gray non-agitation half is pushed high as well, so the model over-calls calm periods as agitation even though the ranking is only partly broken. For others, such as Participant 13, the two halves overlap and the scores barely move with the true state, so the model is closer to indifferent than miscalibrated. The two call for different fixes: over-calling is a specificity and thresholding problem that a per-participant operating point can address, while indifference means the signal for that child is not being captured, which more data or a better sensor fit would have to solve first.

\subsubsection{What Pretraining and Parameter Sharing Buy}
Two results reshape how individualization should be approached here. First, the gap between the from-scratch and pretrained configurations shows that most of the usable signal lives in general-purpose representations pretrained on external corpora, not in structure the model could learn from this cohort alone. Fine-tuning them added little over using them frozen, and in the individual family frozen features were strongest. Second, a single group model with per-participant heads was the best detector, so the individuality of agitation is captured well enough by a shared trunk with a light per-child readout. Together these suggest that scaling to more children is a matter of adding heads to a shared model rather than building a new model for each.

\subsection{Implications for Wearable Design}

Our design targets comfort, acceptability, and coverage for everyday use. The full configuration combines an IMU garment, a physiology device, and a lapel microphone. The garment was selected for autistic participants (e.g., soft seams, predictable fasteners, and fixed sensor pockets) to support tolerance and repeatable placement, guided by a formative wearability study of fabrics, attachment options, and pocket designs (Section~\ref{subsec:wearability_study}).

Our ablation and watch-only results speak to device choice. Audio carried most of the detectable signal, and a watch-only configuration without a microphone stayed near chance ($0.560$ versus $0.699$ for the full model on the same participants). A low-profile physiology device is therefore insufficient for strong detection of individualized agitation in this cohort.

\subsection{Clinical and Translational Implications}
\label{subsec:clinical}
The clinical value of detecting agitation is in the time it buys. Challenging behavior is usually managed reactively, once an episode is already underway, and a signal that flags rising agitation even seconds before it is visible opens a window to change the situation first. Our advance-detection results are modest at longer horizons (Figure~\ref{fig:forecast}), but the response to an early sign is a small change in context and not a clinical procedure, so a brief and reliable alert is actionable. We discuss three settings where such an alert would be meaningful.

The first is therapy, the setting our data came from. A BCBA already watches for a child's individualized early signs and ends the evoking condition when one appears, so a monitor here extends an existing clinical practice. The same score that drives an alert is a continuous record of when agitation arose and how it responded to what the clinician did, and a functional assessment currently has to reconstruct that from memory and hand-coded observation. A BCBA is present and calibrating to that child, so this setting also tolerates the per-participant variation in Figure~\ref{fig:score_sep} that a fixed threshold cannot.

The second is everyday life at home and at school, where most of a child's day happens and where trained observation is least available. Caregivers and teachers come to know a child's signs well, but they cannot watch continuously while handling their other responsibilities. Furthermore, the autonomic component of agitation is not visible to them at all. A wearable that runs through the day extends coverage into the hours outside of clinical observation, and it does so without asking an adult to divide their attention. It would also record escalation where it actually occurs, across the settings and demands that a clinic session can only approximate. That is the context a functional assessment needs and rarely has.

The third setting, and the direction we regard as the goal of this line of work, is for the person to use to alert themselves. Agitation is a state the individual is experiencing. A system that reports only to other people leaves them the subject of monitoring. A wearable that surfaces rising arousal to the person wearing it could support self-regulation directly. It could prompt a break, or a coping strategy the person chose in advance. Over time it could build awareness of a build-up that is hard to notice from the inside. That is what the technology is ultimately for: widening a person's own capacity to act before escalation.

The second and third settings are not reachable with the sensing system we used here. Our full configuration combines an instrumented garment, a physiology device, and a lapel microphone. The microphone is not optional, since the watch-only configuration stayed near chance (Section~\ref{subsec:watch}). Everyday use requires hardware that is miniaturized and robust across a full day of ordinary activity. It also has to be comfortable enough that a child keeps it on without an adult managing it. This work demonstrates that the signal is present and that a shared model can recover it, not that the present system is ready to leave the clinic.

Across all three settings we see this as decision support and not automation. The model outputs a graded score, and what that score should trigger depends on the child and on who is reading it. A threshold that suits one child will over-call for another (Figure~\ref{fig:score_sep}), so the operating point belongs to the person using it. A deployed system would surface a confidence to the clinician, the caregiver, or the individual it serves, and leave the decision with them.

\subsection{Ethical Considerations}

Ethical principles were central in the design and implementation of this study. The data collection protocol sought to prioritize safety, dignity, and comfort by focusing on non-dangerous agitation. Caregivers and participants were informed of their right to pause or withdraw at any time, and sessions were conducted under BCBA oversight with predefined safety plans.

From a modeling perspective, our reliance on pretrained representations was a deliberate choice to reduce the amount of data required per participant, avoiding extensive procedures designed solely to elicit challenging behaviors. The emphasis on personalized, agitation-focused prediction is intended to support proactive, least-restrictive interventions rather than to enable punitive or coercive responses. At the same time, any broader deployment of such systems would require ongoing attention to data privacy, secure storage and transmission of physiological and behavioral data, and transparent communication with families and clinicians about what the system does and does not infer. Continued collaboration with autistic individuals, families, and clinicians will be essential to ensure that future systems remain responsive to the communities they are intended to serve.

\subsection{Limitations and Future Work}
Several limitations frame where this work should go next. The cohort, while larger than prior pilots, is still small and narrow in age, developmental profile, and topography, and performance is bounded by cohort size rather than model capacity. Because heterogeneity in this cohort is individual rather than clustered, scaling the dataset would improve the shared representation without reducing the per-child variation a deployment must absorb. The labels add their own uncertainty, since subtle shifts in posture, expression, or vocal tone are hard to timestamp precisely and blur the boundary between agitation and non-agitation windows near onsets, and how much annotation granularity affects performance is itself worth studying. The advance-detection result relies on an assumption that pre-onset windows already exhibit signs of agitation. Distinguishing genuine early detection from temporal autocorrelation will require continuous intensity ratings or prospective clinician review of model-flagged windows.

Detection was predominantly audio-driven, performing weakest on the facial, postural, and withdrawal signs that microphones and upper-body IMUs capture poorly. Furthermore, the watch-only configuration performed near chance level across the seven children assessed, indicating that identifying sensing modalities better suited for these signs remains an open challenge. The gain from fine-tuning over frozen features was not statistically distinguishable, so the value here is the pretrained representations rather than task adaptation. 

Despite these limitations, this work produces a multimodal machine-learning system for the proactive detection of agitation, using data from wearable sensors and a custom annotation tool. The goal is to give BCBAs valuable intervention time by detecting agitation before it escalates into challenging behavior. Technology of this kind could improve the safety and efficiency of treatment by making escalation visible early enough to act on, whether to an interventionist, a caregiver, or eventually the person themselves (Section~\ref{subsec:clinical}).

\section{Conclusion}
\label{sec:conclusion}

This work addressed the detection of agitation in autistic youth, a setting constrained by data scarcity and highly individualized expression. Rather than train a task-specific model per child, we adapted four foundation models pretrained on movement, audio, photoplethysmography, and electrodermal corpora, projected each to a shared representation, and fused them into a single group model with per-participant heads. This model was then evaluated within a clinically grounded modified PFA/IISCA protocol. Across all 15 participants the model detected clinician-annotated agitation with an AUC of $0.724$ (95\% CI $0.650$--$0.794$), and its detectability declined smoothly as the target was shifted earlier, to $0.608$ at $30$\,s before the annotation. Pretraining was the primary driver of these results. A from-scratch model of the same architecture achieved only 0.58, whereas both frozen and fine-tuned representations yielded better performance. Audio was the dominant modality, and a watch-only configuration stayed near chance. A single shared model was the strongest detector.

These findings establish that individualized agitation is detectable from wearable and audio sensing and that foundation-model transfer captures it with a single shared model. We view this work as a step toward assistive technologies that augment clinical judgment rather than replace it.

\section{Acknowledgement}
We thank the children and families who generously participated in this research. We also thank Gabi Castillo-Martinez, Gabija Zilinskaite, and Adithyan Rajaraman for their essential contributions to completing this study.

\section{Funding}
This work was supported by the National Science Foundation (NSF) grant 2124002.

\bibliographystyle{unsrt}  
\bibliography{references}  

\clearpage
\appendix
\section{Participant-Level Results}

Table~\ref{tab:participant_results} gives the per-participant breakdown behind the cohort averages in Section~\ref{sec:results}. For each child it lists the number of annotated agitation events, the group fine-tuned model's detection AUC with a $95\%$ bootstrap confidence interval, and the best per-child individual model as a point estimate. The two AUC columns are close for most participants, which is the per-child form of the main result that a shared trunk with per-participant heads suffices. The exceptions run in both directions: Participant 15's own model beats the shared one ($0.802$ versus $0.681$), as does Participant 6's ($0.763$ versus $0.683$), while Participant 11's is worse ($0.478$ versus $0.559$). Participants 11 and 13 are the two children not above chance, with Participant 11's interval crossing $0.5$ and Participant 13's falling below it.

\begin{table}[t]
    \centering
    \small
    \setlength{\tabcolsep}{4pt}
    \begin{tabular}{cccc}
    \hline
    Participant & Events & Group AUC [95\% CI] & Individual AUC \\ \hline
    1  & 223 & 0.790 [0.75, 0.83] & 0.806 \\
    2  & 241 & 0.916 [0.89, 0.94] & 0.858 \\
    3  & 366 & 0.895 [0.87, 0.92] & 0.853 \\
    4  & 81  & 0.730 [0.66, 0.79] & 0.637 \\
    5  & 66  & 0.710 [0.64, 0.78] & 0.642 \\
    6  & 333 & 0.683 [0.62, 0.74] & 0.763 \\
    7  & 211 & 0.672 [0.62, 0.72] & 0.636 \\
    8  & 81  & 0.810 [0.76, 0.87] & 0.747 \\
    9  & 80  & 0.859 [0.81, 0.91] & 0.830 \\
    10 & 64  & 0.920 [0.87, 0.96] & 0.894 \\
    11 & 106 & 0.559 [0.49, 0.63]$^{*}$ & 0.478 \\
    12 & 61  & 0.632 [0.54, 0.72] & 0.562 \\
    13 & 58  & 0.394 [0.32, 0.47]$^{*}$ & 0.507 \\
    14 & 196 & 0.607 [0.56, 0.66] & 0.587 \\
    15 & 96  & 0.681 [0.61, 0.74] & 0.802 \\ \hline
    \end{tabular}
    \caption{Per-participant annotated agitation event counts and detection AUC. Group AUC is the shared fine-tuned model with a 95\% bootstrap confidence interval; Individual AUC is the best per-child model (frozen features), a point estimate. $^{*}$Participants 11 and 13 are not above chance: Participant 11's interval crosses $0.5$ and Participant 13's falls entirely below it; all others are above chance.}
    \label{tab:participant_results}
\end{table}

\clearpage
\section{Participant Details}

Table~\ref{tab:participant_summary} describes how agitation and challenging behavior presented for each participant, compiled from clinician input and session review. These profiles are the context behind the per-participant variation discussed in Section~\ref{sec:discussion}: children with discrete, high-energy signs tend to separate well from their calm baseline, while those whose agitation is subtle, postural, or withdrawn tend to sit closer, though this is a tendency with clear exceptions rather than a rule. We include the full descriptions so the per-participant results can be read against the specific behaviors each child showed rather than a single label.

\begin{longtable}{
  P{0.06\textwidth}
  P{0.12\textwidth}
  P{0.06\textwidth}
  P{0.32\textwidth}
  P{0.32\textwidth}
}
\hline
ID & Age (yrs) & Sex & Identified Signs of Agitation & Identified Challenging Behaviors\\
\hline
\endfirsthead

\hline
ID & Age (yrs) & Sex & Identified Signs of Agitation & Identified Challenging Behaviors\\
\hline
\endhead

\hline

1 & 10 yrs 1 mo & M & Facial expression–mean look, squeezing hands/balled-up fist, tugging on shirt, tensing whole body, clicking mouth/odd vocalizations, "You're annoying me," "I'm getting mad/angry," angry vocal tone, turns back or head down, pulling item away, pulling back from space/person, suddenly stops talking & Eloping, physical aggression (pushing, hitting, scratching, biting), property destruction, throwing items \\ 

2 & 11 yrs & F & Frustrated/sad face, "No!", "I don't have to do it", louder/firmer/faster speech, hand grabbing/squeezing, repetitive speech/scripting, moves away, pulls item away & Throwing, crying, hitting, scratching, rare tantrums \\ 

3 & 11 yrs & M & Louder vocalization/screaming, "First", angry face, slapping on iPad/surfaces, pushing materials away while saying "No", pulling away iPad & Throwing items, hitting, kicking, property destruction (throwing iPad) \\ 

4 & 3 yrs 5 mo & F & Crying/whining, throwing materials, trying to grab removed object, clearing items, turning away, tug-of-war with item, grabbing/squeezing to stop interference, suddenly falling to knees, throwing materials, grabbing removed items & Self-injury (head-banging) \\ 

5 & 3 yrs 5 mo & M & Vocal stim, throwing arms down, walking away/in circles, reaching/vocalizing after denial, sad/distressed face, quick dip to side, ball-up/squeeze body, stomping, turning/hiding item, kicking/stomping feet & Self-injury (head hitting), squeezing others' arms, crying/meltdowns \\ 

6 & 13 yrs 1 mo & M & Facial drop, jaw clench, grabbing adult's hand, stern "stop it/no", pushing hand away, pulling item away, scooching away, forceful toe-jumping, vocal disruptions & Self-injury (head hitting) \\ 

7 & 7 yrs 4 mo & M & "No, no, no" (fast, hands up), "Wait, wait, wait", balled fists, widened eyes, walking into someone's face, turning/twisting away, head/chin tics, pushing hand away & Eloping, kicking/shoving peers, yelling, head hitting (when sick) \\ 

8 & 5 yrs 10 mo & F & Rapid breathing, body/muscle tensing, angry protesting tone, crying, eyebrows down, lips pushed together/down, folding arms/legs, turning back & Elopement, hitting others, property destruction \\ 

9 & 3 yrs 7 mo & M & Screaming, loud noises, squinting/covering eyes, covering ears, "No, no, no", growling, pushing away, pulling item away/moving away, stomping/kicking, throwing item while looking for reaction & Hair pulling, punching, biting, kicking, pushing \\ 

10 & 17 yrs 5 mo & M & Angry talking, banging on table, knocking head on fist, head down on arm, mumbling angrily, angry eyebrows, fists balled, knocking chair over, "You better not…", "You do too much…", insults with expletives, side-eye, hard breathing & Punching holes in wall, verbal aggression, physical aggression \\ 

11 & 5 yrs 10 mo & F & High-pitched whine, stomping feet, crying, snatching item, walking away/pushing person out of her space, crossing arms/pouting, squinting/scrunched eyebrows, tapping with heel of palm, shaking arm of person, side-eye & Self-injury (head hitting, flopping to ground) \\

12 & 5 yrs 9 mo & M & High-pitched whine, pushing item/hand away, grimacing with teeth shown, tensing arms, single stomp, turning/moving away from adult/out of range & Throwing self onto floor/wall (low intensity), fussing \\

13 & 6 yrs 7 mo & M & Crying, loud grunts, eye contact then light slap, "Oh no!", withholding/protecting items, kicking doors, laying down and kicking, throwing items & Self-injury (chin hitting, head smacking), grabbing ears, clapping intensely when upset \\

14 & 10 yrs 8 mo & M & Yelling, "You need to go to Siberia", stomping, accusations ("Why did you hit me?"), loud/huffy breathing, fists balled, muttering, angry face, side-eye/eye rolling, growling, tensed shoulders, lunging/charging, foot tapping, snapping, kicking items, knocking items off table, protecting iPad, "No more questions please", "You want to replace me" & Physical aggression, property destruction \\

15 & 3 yrs 4 mo & M & Head down, covering ears, cutting off eye contact, pushing hand away, walking away, isolating self, facial grimace, "I need some space", grunting & Elopement, self-injury (throwing body), pinching, biting \\ \hline

\caption{Individualized agitation and challenging behavior descriptions for each participant.}
\label{tab:participant_summary}
\end{longtable}

\end{document}